\documentclass[reqno,draft
]{amsproc}
\usepackage{
amssymb}
\include{diagrams}
\allowdisplaybreaks[4]
\diagramstyle[small,scriptlabels,centerdisplay,midhshaft,nohug]
\newarrow{TeXto}----{->}
\newcommand{\rto}{\rTeXto}
\newcommand{\lto}{\lTeXto}
\newcommand{\uto}{\uTeXto}

\newcommand{\dto}{\dTeXto}

\newcommand{\rdto}{\rdTeXto}
\newcommand{\luto}{\luTeXto}
\newcommand{\ldto}{\ldTeXto}

\newcommand{\cprime}{\/{\mathsurround=0pt$'$}}
\newcommand{\pinner}{\mathbin{\mathchoice
   {\hbox{\vrule width0.6em depth0pt height0.4pt
   \vrule width0.4pt depth0pt height0.8ex}}
   {\hbox{\vrule width0.6em depth0pt height0.4pt
   \vrule width0.4pt depth0pt height0.8ex}}
   {\hbox{\kern0.14em
   \vrule width0.48em depth0pt height0.4pt
   \vrule width0.4pt depth0pt height0.6ex\kern0.14em}}
   {\hbox{\kern0.1em
   \vrule width0.39em depth0pt height0.4pt
   \vrule width0.4pt depth0pt height0.5ex\kern0.1em}}}}

\DeclareMathOperator{\sym}{sym}
\DeclareMathOperator{\im}{im}
\DeclareMathOperator{\id}{id}

\newcommand{\BBR}{\mathbb{R}}
\newcommand{\CB}{\mathcal{B}}
\newcommand{\CC}{\mathcal{C}}
\newcommand{\CE}{\mathcal{E}}

\newcommand{\CU}{\mathcal{U}}
\newcommand{\Dr}{\mathrm{D}}
\newcommand{\Ld}{\mathrm{L}}
\newcommand{\ip}{\mathrm{i}}

\newcommand{\sbs}{\subset}
\newcommand{\wg}{\wedge}
\newcommand{\ot}{\otimes}

\newcommand{\De}{\Delta}
\newcommand{\La}{\Lambda}
\newcommand{\la}{\lambda}
\newcommand{\na}{\nabla}
\newcommand{\Om}{\Omega}
\newcommand{\om}{\omega}
\newcommand{\vf}{\varphi}

\newcommand{\h}{-\hspace{0pt}}

\newcommand{\Ei}{\CE^\infty}
\newcommand{\Ci}{C^\infty}
\newcommand{\Ji}{J^\infty}

\def\ldb{{\rm [\![}}
\def\rdb{{\rm ]\!]}}
\newcommand{\fnij}[2]{\ldb{#1},{#2}\rdb}

\newtheorem{theorem}{Theorem}
\newtheorem{proposition}{Proposition}
\newtheorem{lemma}{Lemma}

\theoremstyle{definition}

\newtheorem{example}{Example}

\theoremstyle{remark}
\newtheorem{remark}{Remark}

\begin{document}
\title[On one-parametric families of Backlund transformations]{On one-parametric families of B\"{a}cklund transformations}
\author{Sergei Igonin}
\author[Joseph Krasilshchik]{Joseph Krasil{\cprime}shchik}
\thanks{Partially supported by the INTAS grant 96-0793.}
\address{The Diffiety Institute, Independent University of Moscow\protect\newline
Correspondence to:
1st Tverskoy-Yamskoy per. 14, Apt. 45, 125047 Moscow, Russia}
\email{igonin@mccme.ru, josephk@online.ru}
\begin{abstract}
In the context of the cohomological deformation theory, infinitesimal
description of one-parametric families of B{\"a}cklund transformations of
special type including classical examples is given. It is shown that any
family of such a kind evolves in the direction of a nonlocal symmetry
shadow in the sense of \cite{KV-trends}.
\end{abstract}
\maketitle

\section*{Introduction}
The role of B{\"a}cklund transformations in constructing exact solutions
of nonlinear partial differential equations is well known, see \cite{Dodd}
and relevant references therein, for example. A general scheme is illustrated by
classical works by B{\"a}cklund and Bianchi. Namely, for the sine\h Gordon
equation
\begin{equation}\label{eq:s-G}
u_{xy}=\sin u
\end{equation}
B{\"a}cklund constructed a system of differential relations $\CB(u,v;\la)=0$
depending on a real parameter $\la\in\BBR$ and satisfying the following
property: if $u=u(x,y)$ is a solution of \eqref{eq:s-G}, then $v$ is a
solution of the same equation and vice versa. Using this result, Bianchi
showed that if a known solution $u_0$ is given and solutions $u_1$, $u_2$
satisfy the relations $\CB(u_0,u_i;\la_i)=0$, $i=1,2$, then there exists a
solution $u_{12}$ which satisfies $\CB(u_1,u_{12};\la_2)=0$,
$\CB(u_2,u_{12};\la_1)=0$ and is expressed in terms of $u_0$, $u_1$, $u_2$ in
terms of relatively simple equalities. This is the so-called \emph{Bianchi
permutability theorem}, or \emph{nonlinear superposition principle}. This
scheme was successfully applied to many other ``integrable'' equations.

Quite naturally, a general problem arises: given an arbitrary PDE $\CE$,
when are we able to implement a similar construction? This question is
closely related to another problem of a great importance in the theory
of integrable systems, the problem of insertion of a nontrivial ``spectral
parameter'' to the initial equation. In this paper, we mainly deal with the
first problem referring the reader to the yet unpublished work by
M.~Marvan \cite{Mar-unp}, where the second problem is analyzed
(see also Remark \ref{rem:triv0} in Section \ref{sec:2}).

Our approach to solution lies in the framework of the geometrical theory
of nonlinear PDE, and the first section of the paper contains a brief
introduction to this theory, including its nonlocal aspects (the theory
of coverings), see \cite{KLV,KV-book,KV-trends}. The second section deals
with cohomological invariants of nonlinear PDE naturally associated to the
equation structure. Our main concern here is the relation between this
cohomology theory and deformations of the structure \cite{K-backgr,K-newinv,
KrasVer}. In the third section, we give a geometrical definition of
B{\"a}cklund transformations and using cohomological techniques prove the
main result of the paper describing infinitesimal part of one-parameter
families of B{\"a}cklund transformations.

\section{Equations and coverings}\label{sec:1}
Let us recall basic facts from the geometry of nonlinear PDE,
\cite{KLV,KV-book}.

Consider a smooth manifold $M$, $\dim M=n$, and a locally trivial smooth
vector bundle $\pi\colon E\to M$. Denote by $\pi_k\colon J^k(\pi)\to M$,
$k=0,1,\dots,\infty$, the corresponding bundles of jets. A \emph{differential
equation} of order $k$, $k<\infty$, in the bundle $\pi$ is a smooth
submanifold $\CE\sbs J^k(\pi)$. To any equation $\CE$ there corresponds a
series of its \emph{prolongations} $\CE^s\sbs J^{k+s}(\pi)$ and the
\emph{infinite prolongation} $\Ei\sbs\Ji(\pi)$. We consider below
\emph{formally integrable} equations, which means that all $\CE^s$ are smooth
manifolds and the natural projection $\pi_\CE=\pi_\infty|_{\Ei}\colon\Ei\to
M$ is a smooth bundle. For any $s>0$ there also exist natural bundles
\begin{equation}\label{eq:proj}
\Ei\xrightarrow{\pi_{\CE,s}}\CE^s\xrightarrow{\pi_{\CE,s,s-1}}\CE^{s-1}
\xrightarrow{\pi_{\CE,s-1}}M
\end{equation}
whose composition equals $\pi_\CE$.
The space $\Ji(\pi)$ is endowed with an integrable
distribution\footnote{Integrability in this context means that $\CC \Dr(\pi)$
satisfies the Frobenius condition: $[\CC \Dr(\pi),\CC \Dr(\pi)]\sbs\CC \Dr(\pi)$.}
denoted by $\CC \Dr(\pi)$. Namely, any point $\theta\in\Ji(\pi)$ is, by
definition, represented in the form $[f]_x^\infty$, $x=\pi_\infty(\theta)\in
M$, where $f$ is a (local) section of $\pi$ such that the graph $M_f^\infty$
of its infinite jet passes through $\theta$ while $[f]_x^\infty$ is the class
of (local) sections $f'$ satisfying the condition
\[
M_{f'}^\infty\text{ is tangent to }M_f^\infty\text{ at }\theta\text{
with infinite order}.
\]
Then the tangent plane $T_\theta M_f^\infty$ is independent of the choice of
$f$ and we set $\CC \Dr(\pi)_\theta=T_\theta M_f^\infty$. The distribution
$\CC \Dr(\pi)$ is $n$\h dimensional and is called the \emph{Cartan
distribution} on $\Ji(\pi)$. Since, by construction, all planes of the
Cartan distribution are horizontal (with respect to $\pi_\infty$) and $n$\h
dimensional, a connection $\CC\colon \Dr(M)\to \Dr(\pi)$ is determined, where
$\Dr(M)$ and $\Dr(\pi)$ denote the modules of vector fields on $M$ and $\Ji(\pi)$
respectively. This connection is flat and is called the \emph{Cartan
connection}.

\begin{remark}\label{rem:C}
In fact, the bundle $\pi_\infty$ possesses a stronger structure than just a
flat connection. Namely, for any vector bundles $\xi$ and $\eta$ over $M$ and
a linear differential operator $\De$ acting from $\xi$ to $\eta$, a linear
differential operator $\CC\De$ acting from the pullback $\pi_\infty^*(\xi)$ to
$\pi_\infty^*(\eta)$ is defined in a natural way. The correspondence $\De
\mapsto\CC\De$ is linear, preserves composition, and the Cartan connection is
its particular case.
\end{remark}

Both the Cartan distribution and the Cartan connection are restricted to
the spaces $\Ei$ and bundles $\pi_\CE$ respectively. The corresponding
objects are denoted by $\CC \Dr(\Ei)$ and $\CC=\CC_\CE\colon \Dr(M)\to \Dr(\Ei)$,
where $\Dr(\Ei)$ is the module of vector fields on $\Ei$. The characteristic
property of the Cartan distribution $\CC \Dr(\Ei)$ on $\Ei$ is that its
maximal integral manifolds are solutions of the equation $\CE$ and vice
versa. The connection form $U_\CE\in \Dr(\La^1(\Ei))$ of the connection
$\CC_\CE$ is called the \emph{structural element} of the equation $\CE$.
Here $\Dr(\La^1(\Ei))$ denotes the module of derivations $\Ci(\Ei)\to
\La^1(\Ei)$ with the values in the module of one-forms on $\Ei$.

Denote by $\Dr_\CC(\Ei)$ the module
\[
\Dr_\CC(\Ei)=\{\,X\in \Dr(\Ei)\mid [X,\CC \Dr(\Ei)]\sbs\CC \Dr(\Ei)\,\}.
\]
Then $\Dr_\CC(\Ei)$ is a Lie algebra with respect to commutator of vector
fields and due to integrability of the Cartan distribution $\CC \Dr(\Ei)$ is
its ideal. The quotient Lie algebra $\sym\CE=\Dr_\CC(\Ei)/\CC \Dr(\Ei)$ is
called the \emph{algebra of \textup{(}higher\textup{)} symmetries} of the
equation $\CE$. Denote by $\Dr^v(\Ei)$ the module of $\pi_\CE$\h vertical
vector fields on $\Ei$. Then in any coset $X\bmod\CC \Dr(\Ei)\in\sym\CE$ there
exists a unique vertical element and this element is called a
\emph{\textup{(}higher\textup{)} symmetry} of $\CE$.

\begin{remark}\label{rem:2}
It may so happen that a coset $X\bmod\CC \Dr(\Ei)$ contains a representative
$X'$ which is protectable to a vector field $X'_s$ on $\CE^s$ by
$\pi_{\CE,s}$ for some $s<\infty$ (see \eqref{eq:proj}). Then it can be
shown that $X'$ is protectable to all $\CE^s$ and $(\pi_{\CE,s,s-1})_*X'_s=
X'_{s-1}$. In this case, $X'$ is called a \emph{classical
\textup{(}infinitesimal\textup{)} symmetry} of $\CE$ an possesses
trajectories in $\Ei$. The corresponding diffeomorphisms preserve solutions
of $\CE$ and are called \emph{finite symmetries}.
\end{remark}

We now pass to a generalization of the above described geometrical theory,
the theory of coverings \cite{KV-trends}. Let $\tau\colon W\to\Ei$ be a
smooth fiber bundle, the manifold $W$ being equipped with an integrable
distribution $\CC \Dr_\tau(W)=\CC \Dr(W)\sbs \Dr(W)$ of dimension $n=\dim M$. Then
$\tau$ is called a \emph{covering} over $\CE$ (or over $\Ei$), if for any
point $\theta\in W$ one has $\tau_*(\CC \Dr(W)_\theta)=
\CC \Dr(\Ei)_{\tau(\theta)}$. Equivalently, a covering structure in the bundle
$\tau$ is determined by a flat connection $\CC_\tau\colon \Dr(M)\to \Dr(W)$
satisfying $\tau_*\circ\CC_\tau=\CC_\CE$. Let $U_\tau\in \Dr(\La^1(W))$ be the
corresponding connection form. We call it the \emph{structural element} of
the covering $\tau$.

\begin{example}[see \cite{Mar-look}]\label{ex:vert}
Let $\CE\sbs J^k(\pi)$ be an equation. Consider the tangent bundle $T\Ei\to
\Ei$ and the subbundle $\pi_\CE^v\colon T^v\pi_\CE\to\Ei$, where
$T^v\pi_\CE$ consists of $\pi_\CE$\h vertical vectors. Hence, the module of
sections for $\pi_\CE^v$ consists of $\pi_\CE$\h vertical vector fields on
$\Ei$.

Then $\pi_\CE^v$ carries a natural covering structure. Namely, for any vector
field $X\in \Dr(M)$ and a vertical vector field $Y$ we set
$[\CC_{\tau^v}(X),Y]=[\CC_\CE(X),Y]\pinner U_\CE$, where $U_\CE$ is the
structural element of the equation $\CE$. It is easily seen that the
connection $\CC_{\tau^v}$ is well defined in such a way and projects to
the connection $\CC_\CE$.
\end{example}

Given two coverings $\tau_i\colon W_i\to\Ei$, $i=1,2$, we say that a smooth
mapping $F\colon W_1\to W_2$ is a \emph{morphism} of $\tau_1$ to $\tau_2$,
if
\begin{enumerate}
\item[(i)] $F$ is a morphism of fiber bundles,
\item[(ii)] $F_*$ takes the distribution $\CC \Dr_{\tau_1}(W_1)$ to
$\CC \Dr_{\tau_2}(W_2)$ (equivalently, $F_*\circ
\CC_{\tau_1}=\CC_{\tau_2}$).
\end{enumerate}
A morphism $F$ is said to be an \emph{equivalence}, if it is a diffeomorphism.

Similar to the case of infinitely prolonged equations, we can define the
Lie algebra $\Dr_{\CC_\tau}(W)$ such that $\CC \Dr_{\tau}(W)$ is it its ideal and
introduce the algebra of \emph{nonlocal $\tau$\h symmetries} as the quotient
$\sym_\tau\CE=\Dr_{\CC_\tau}(W)/\CC \Dr_{\tau}(W)$. Again, in any coset
$X\bmod\CC_\tau \Dr(W)\in\sym_\tau\CE$ there exists a unique
$(\pi_\CE\circ\tau)$\h vertical representative and it is called
a \emph{nonlocal $\tau$\h symmetry} of the equation $\CE$.

Obviously, one can introduce the notion of a covering over covering, etc. In
particular, the subbundle $\pi_\CE^v\colon T^v\pi_\CE\to\Ei$ of
$(\pi_\CE\circ\tau)$\h vertical vectors (cf.~Example~\ref{ex:vert}) is a
covering over $\CE$ while the intermediate projection
$\tau^v\colon T^v\tau\to W$ is a covering over $W$.
Note also that the correspondence $\tau\Rightarrow \tau^v$
determines a covariant functor in the category of coverings.

We shall now reinterpret the concepts of a symmetry and nonlocal symmetry
using the results of \cite{Mar-look}. Namely, one has
\begin{proposition}
Let $\CE$ be an equation and $\tau\colon W\to\Ei$ be a covering over it.
Then\textup{:}
\begin{enumerate}
\item
There is a one-to-one correspondence between symmetries of $\CE$ and sections
$\vf\colon\Ei\to T^v\pi_\CE$ of the bundle $\pi_\CE^v\colon T^v\pi_\CE\to\Ei$
such that $\vf_*$ takes the Cartan distribution on $\Ei$ to that on
$T^v\pi_\CE$.
\item
There is a one-to-one correspondence between nonlocal $\tau$\h symmetries of
$\CE$ and sections $\psi$ of the bundle $(\pi_\CE\circ\tau)^v\colon
T^v(\pi_\CE\circ\tau)\to W$ such that $\psi_*$ takes the Cartan distribution
on $W$ to that on $T^v\tau$.
\end{enumerate}
\end{proposition}

Let us say that a mapping $s\colon W\to T^v\pi_\CE$ is a \emph{$\tau$\h
shadow} of a nonlocal symmetry (cf.~\cite{KV-trends}), if $\pi_\CE^v\circ s=
\tau$ and $s_*$ preserves the Cartan distribution.

\begin{example}\label{sym-shad}
Any symmetry $\vf$ considered as a section $\vf\colon\Ei\to T^v\pi_\CE$
determines a shadow $\vf\circ\tau$.
\end{example}

\begin{proposition}[The shadow reconstruction theorem]\label{pr:shad-rec}
For an arbitrary covering $\tau\colon W\to\Ei$ and a $\tau$\h shadow $s\colon W\to
T^v\pi_\CE$ there exists a covering $\tau'\colon W'\to W$ and a nonlocal
$\tau\circ\tau'$\h symmetry $s'\colon W'\to T^v(\pi_\CE\circ\tau\circ\tau')$
such that the diagram
\begin{diagram}[LaTeXeqno]\label{diag:shad}
 &              &T^v\pi_\CE&\lto^{(\tau\circ\tau')_*}& & &T^v(\pi_\CE\circ\tau\circ\tau')\\
 &              &\dto^{\pi_\CE^v}&\luto^s    & & &\dto^{(\pi_\CE\circ\tau\circ\tau')^v}\uto_{s'}\\
M&\lto^{\pi_\CE}&\Ei             &\lto^\tau&W&\lto^{\tau'}&W'\\
\end{diagram}
is commutative. In other words\textup{,} any shadow can be reconstructed
up to a nonlocal symmetry in some new covering.
\end{proposition}
\begin{proof}
Consider the following commutative diagram:
\begin{diagram}[width=3em]
 &    &T^v\pi_\CE&\lto &\ T^v(\tau\circ\pi_\CE)\ &\lto &\ T^v(\tau^v\circ\tau\circ\pi_\CE)\ &\lto &\dotsb   &    &\\
 &    &\dto^{\pi_\CE^v}      &\luto^s&\dto_{(\tau\circ\pi_\CE)^v}                 &\luto^{s_*}&\dto_{(\tau^v\circ\tau\circ\pi_\CE)^v}                            &\luto^{(s_*)_*}&         &    &\\
M&\lto_{\pi_\CE}&\Ei       &\lto_{\tau} &W                    &\lto_{\tau^v} &T^v\tau                         &\lto_{(\tau^v)^v} &T^v\tau^v&\lto&\dotsb\\
\end{diagram}
and let us set $\tau_0=\tau$, $\tau_{i+1}=\tau_i^v$, $W_0=W$, $W_i=T^v\tau_i$,
$s_0=s$, $s_{i+1}=(s_i)_*$, where $s_*=ds$. Then the above diagram is infinitely continued to
the left, while by setting $\bar{\tau}_i=\tau_1\circ\dots\circ\tau_i$ and passing
to the inverse limit, we obtain Diagram \ref{diag:shad} with $\tau'=
\bar{\tau}_\infty$, $s'=s_\infty$, and $W'=W_\infty$.
\end{proof}

\section{$\CC$-complex and deformations}\label{sec:2}
We now pass to describe a cohomological theory naturally related to covering
structures and supplying their important invariants, cf.~\cite{K-newinv}.

Let $W$ be a smooth manifold and $\Dr(\La^i(W))$ denote the $\Ci(W)$\h module
of $\La^i(W)$\h valued derivations $\Ci(W)\to\La^i(W)$. For any element
$\Om\in\La^i(W)$ one can define the \emph{inner product} operation
\[
\ip_\Om\colon\La^j(W)\to\La^{i+j-1}(W),
\]
also denoted by $\Om\pinner\rho$, $\rho\in\La^*(W)$, and the \emph{Lie
derivative} along $\Om$:
\[
\Ld_\Om=[\ip_\Om,d]\colon\La^j(W)\to\La^{i+j}(W),
\]
where $[\ip_\Om,d]$ denotes the \emph{graded} commutator.

Then for any two elements $\Om,\Theta\in \Dr(\La^*(W))$ we can introduce their
\emph{Fr\"{o}licher--Nijenhuis bracket} by setting
\[
\fnij{\Om}{\Theta}(f)=\Ld_\Om(\Theta(f))-(-1)^{ij}\Ld_\Theta(\Om(f)),
\]
where $f\in\Ci(W)$ and $i,j$ are degrees of $\Om$ and $\Theta$
respectively\footnote{We say that $i$ is the degree of $\Om$, if $\Om\in
\Dr(\La^i(W))$.}.

\begin{remark}
In the sequel, we shall also need the following facts.
\begin{enumerate}
\item
In the case, when $W$ is a finite\h dimensional manifold, one has an
isomorphism
$\Dr(\La^*(W))\simeq\La^*(W)\ot \Dr(W)$ and thus any derivation $\Om\in
\Dr(\La^*(W))$ is representable as a finite sum of elements of the form
\begin{equation}\label{eq:decomp}
\Om=\om\ot X,
\end{equation}
where $\om\in\La^*(W)$ and $X\in \Dr(W)$. For an arbitrary $W$, an
embedding $\La^*(W)\ot \Dr(W)\sbs \Dr(\La^*(W))$ is defined by $(\om\ot X)f=
X(f)\om$.
\item
For elements of the form \eqref{eq:decomp}, one has
\[
(\om\ot X)\pinner\rho=\om\wg(X\pinner\rho),\quad
\Ld_{\om\ot X}\rho=\om\wg \Ld_X\rho+(-1)^id\om\wg(X\pinner\rho)
\]
and
\begin{align*}
\fnij{\om\ot X}{\theta\ot Y}&=\om\wg\theta\ot[X,Y]+\om\wg \Ld_x(\theta)\ot Y
+(-1)^id\om\wg(X\pinner\theta)\ot Y\\
&-(-1)^{ij}\theta\wg \Ld_Y(\om)\ot X-(-1)^{(i+1)j}d\theta\wg(Y\pinner\om)\ot X,
\end{align*}
where $X,Y\in \Dr(W)$, $\om\in\La^i(W)$, $\theta\in\La^j(W)$.
\item
Note also that another two operations are defined on elements of the module
$\Om\in \Dr(\La^*(W))$: we can multiply elements of $\Dr(\La^*(W))$ by forms
$\rho\in\La^*(W)$ and for $\Om=\om\ot X$ one has $\rho\wg\Om=(\rho\wg\om)\ot
X$. In addition, we can insert elements of $\Dr(\La^*(W))$ into each other;
in representation \eqref{eq:decomp} this operation is represented as
\[
(\om\ot X)\pinner(\theta\ot Y)=\om\wg(X\pinner\theta)\ot Y.
\]
\end{enumerate}
\end{remark}

The basic properties of the above introduced operations are formulated in
\begin{proposition}[see \cite{K-backgr}]\label{pr:prop}
Let $\Om\in \Dr(\La^i(W))$, $\Theta\in \Dr(\La^j(W))$, $\rho\in\La^k(W)$, and
$\eta\in\La^l(W)$. Then\textup{:}
\begin{itemize}
\item[(i)]
$\ip_\Om(\rho\wg\eta)=\ip_\Om(\rho)\wg\eta+(-1)^{(i-1)k}\rho\wg
\ip_\Om(\eta)$\textup{;}
\item[(ii)]
$\ip_\Om(\rho\wg\Theta)=\ip_\Om(\rho)\wg\Theta+(-1)^{(i-1)k}\rho\wg
\ip_\Om(\Theta)$\textup{;}
\item[(iii)]
$[\ip_\Om,\ip_\Theta]=\ip_{\fnij{\Om}{\Theta}^{\mathrm{rn}}}$\textup{,} where
\[
\fnij{\Om}{\Theta}^{\mathrm{rn}}=\ip_\Om\Theta-(-1)^{(i-1)(j-1)}\ip_\Theta\Om
\]
is the \emph{Richardson--Nijenhuis bracket} of $\Om$ and $\Theta$\textup{;}
\item[(iv)]
$\Ld_\Om(\rho\wg\eta)=\Ld_\Om(\rho)\wg\eta+(-1)^{ik}\om\wg \Ld_\Om(\eta)$\textup{;}
\item[(v)]
$\Ld_{\rho\wg\Om}=\rho\wg \Ld_\Om+(-1)^{i+k}d\om\wg \ip_\Om$\textup{;}
\item[(vi)]
$[\Ld_\Om,d]=0$\textup{;}
\item[(vii)]
$[\Ld_\Om,\Ld_\Theta]=\Ld_{\fnij{\Om}{\Theta}}$\textup{;}
\item[(viii)]
$\fnij{\Om}{\Theta}+(-1)^{ij}\fnij{\Theta}{\Om}=0$\textup{;}
\item[(ix)]
$\fnij{\Om}{\fnij{\Theta}{\Xi}}=\fnij{\fnij{\Om}{\Theta}}{\Xi}+
(-1)^{ij}\fnij{\Om}{\fnij{\Theta}{\Xi}}$\textup{,} where
$\Xi\in \Dr(\La^m(W))$\textup{;}
\item[(x)]
$[\Ld_\Om,\ip_\Theta]=\ip_{\fnij{\Om}{\Theta}}
-(-1)^{i(j+1)}\Ld_{\Theta\pinner\Om}$\textup{;}
\item[(xi)]
$\Xi\pinner\fnij{\Om}{\Theta}=
\fnij{\Xi\pinner\Om}{\Theta}+(-1)^{i(m+1)}\fnij{\Om}{\Xi\pinner\Theta}
+(-1)^i\fnij{\Xi}{\Om}\pinner\Theta\\
-(-1)^{(i+1)j}\fnij{\Xi}{\Theta}\pinner\Om$\textup{;}
\item[(xii)]
$\fnij{\Om}{\rho\wg\Theta}=(\Ld_\Om\rho)\wg\Theta
-(-1)^{(i+1)(j+k)}d\rho\wg \ip_\Theta\Om
+(-1)^{ik}\rho\wg\fnij{\Om}{\Theta}$.
\end{itemize}
\end{proposition}

In particular, from Proposition \ref{pr:prop}~(ix) it follows that for
$\Om\in \Dr(\La^1(W))$ satisfying the \emph{integrability property}
\begin{equation}\label{eq:integr}
\fnij{\Om}{\Om}=0
\end{equation}
the mapping
\[
\partial_\Om=\fnij{\Om}{\cdot}\colon \Dr(\La^i(W))\to \Dr(\La^{i+1}(W))
\]
is a differential, i.e., $\partial_\Om\circ\partial_\Om=0$, and thus we
obtain the complex
\begin{equation}\label{eq:compl}
0\to \Dr(W)\to\dots\to \Dr(\La^i(W))\xrightarrow{\partial_\Om}\Dr(\La^{i+1}(W))\to
\dotsb
\end{equation}

Assume now that the manifold $W$ is fibered by $\xi\colon W\to M$ and a
connection $\na$ is given in the bundle $\xi$. Then the following fact
is valid:
\begin{proposition}[cf.~\cite{K-flat}]
\[
\fnij{U_\na}{U_\na}=2R_\na,
\]
where $U_\na$ is the connection form and $R_\na$ is the curvature.
\end{proposition}

Consequently, if $\na$ is a flat connection, i.e., $R_\na=0$, then $\Om=
U_\na$ enjoys the integrability property \eqref{eq:integr} and to any
flat connection a complex of the form \eqref{eq:compl} corresponds. In this
case, we shall use the notation $\partial_\Om=\partial_\na$.

Now, we pass to the case of our main interest: let $\xi$ be the composition
$W\xrightarrow{\tau}\Ei\xrightarrow{\pi_\CE}M$, $\tau$ being a covering over
$\CE$, and $\na$ be the Cartan connection $\CC_\tau$ associated to the
covering structure. We include in consideration the case $W=\Ei$, $\tau=\id$,
and $\CC_\tau=\CC_\CE$. Let us restrict complex \eqref{eq:compl} to
\emph{vertical} derivations, i.e., to derivations
\[
\Dr^v(\La^i(W))=\{\,\Om\in \Dr(\La^i(W))\mid\Om(f)=0,\forall f\in\Ci(M)\,\}.
\]
By construction, $U_\tau$ (or $U_\CE$) lies in $\Dr^v(\La^1(W))$ (resp., in
$\Dr^v(\La^1(\Ei))$), while from the definition of the
Fr\"{o}licher--Nijenhuis bracket it follows that the differential in
\eqref{eq:compl} preserves vertical derivations. The vertical part of
\eqref{eq:compl} will be denoted by
\begin{equation}\label{eq:compl-v}
0\to \Dr^v(W)\to\dots\to \Dr^v(\La^i(W))\xrightarrow{\partial_\tau}
\Dr^v(\La^{i+1}(W))\to\dotsb
\end{equation}
or
\begin{equation}\label{eq:compl-ve}
0\to \Dr^v(\Ei)\to\dots\to \Dr^v(\La^i(\Ei))\xrightarrow{\partial_\CE}
\Dr^v(\La^{i+1}(\Ei))\to\dotsb,
\end{equation}
when the equation is considered as is. The cohomology of \eqref{eq:compl-v}
(resp., of \eqref{eq:compl-ve}) is denoted by $H_\CC(\CE;\tau)$ (resp., by
$H_\CC(\CE)$) and is called the \emph{$\CC$\h cohomology} of the covering
$\tau$ (resp., of the equation $\CE$). The following fundamental result is
valid:
\begin{theorem}[cf.~\cite{K-newinv}]\label{thm:deform}
Let $\CE\sbs J^k(\pi)$ be a formally integrable equation and $\tau\colon
W\to \Ei$ be a covering over $\CE$. Then\textup{:}
\begin{enumerate}
\item
The module $H_\CC^0(\CE;\tau)$ is isomorphic to the Lie algebra $\sym_\tau
\CE$ of nonlocal $\tau$\h symmetries \textup{(}resp.\textup{,}
$H_\CC^0(\CE;\tau)$ is isomorphic to $\sym\CE$\textup{)}.
\item
The module $H_\CC^1(\CE;\tau)$ is identified with equivalence classes of
nontrivial infinitesimal deformations of the covering structure $U_\tau$
\textup{(}resp.\textup{,} of the equation structure $U_\CE$\textup{)}.
\item
The module $H_\CC^2(\CE;\tau)$ consists of obstructions to prolongation of
infinitesimal deformations up to formal ones.
\end{enumerate}
\end{theorem}

\begin{remark}\label{rem:triv0}
Of course, if $U_\la$ is a deformation of the equation structure, the
condition that $d U_\la/d\la|_{\la=0}$ lies in $\ker\partial_\CE$ is not
sufficient for this deformation to be trivial. Nevertheless, the following
fact is obviously valid:
\begin{proposition}\label{prop:triv}
Let $U_\la$ be a smooth deformation of the equation structure $U=U_\CE$
satisfying the condition
\begin{equation}\label{eq:triv}
\frac{dU_\la}{d\la}=\fnij{X_\la}{U_\la},
\end{equation}
where $X_\la$ is a smooth vector field on $\Ei$ for any $\la$ with smooth
dependence on $\la$. Then $U_\la$ is uniquely defined by \eqref{eq:triv}
and is of the form
\[
U_\la=\exp(\int_0^\la X_\mu\,d\mu)U,
\]
where the left- and right hand sides are understood as formal series.
In this sense, $U_\la$ is formally trivial.
\end{proposition}
\end{remark}

Let us now consider the mapping $\Ld_{U_\tau}\colon\La^i(W)\to\La^{i+1}(W)$
and denote it by $d_\CC$. Since the element $U_\tau$ is integrable, one has
the identity $d_\CC\circ d_\CC=0$. We call $d_\CC$ the \emph{vertical}, or
\emph{Cartan differential} associated to the covering structure. Due to
Proposition \ref{pr:prop}~(vi), $[d,d_\CC]=0$ and consequently the mapping
$d_h=d-d_\CC$ is also a differential and $[d_h,d_\CC]=0$. The differential
$d_h$ is called the \emph{horizontal differential}, while the pair
$(d_h,d_\CC)$ forms a bicomplex with the total differential $d$. The
corresponding spectral sequence coincides with the Vinogradov $\CC$\h
spectral sequence for the covering $\tau$, \cite{Vin-CSP}.

Denote by $\La_h^1(W)$ the submodule in $\La^1(W)$ spanned by $\im d_h$ and
by $\CC^1\La(W)$ the submodule generated by $\im d_\CC$. Then the direct sum
decomposition $\La^1(W)=\La^1_h(W)\oplus\CC\La^1(W)$ takes place and
generates the decomposition
\[
\La^i(W)=\bigoplus_{p+q=i}\CC^p\La(W)\ot\La^q_h(W)=
\bigoplus_{p+q=i}\La^{p,q}(W),
\]
where
\[
\CC^p\La(W)=\underbrace{\CC^1\La(W)\wg\dots\wg\CC^1\La(W)}_{p\
\mathrm{times}},\quad
\La^q_h(W)=\underbrace{\La^1_h(W)\wg\dots\wg\La^1_h(W)}_{q\
\mathrm{times}}.
\]
Then $d_\CC\colon\La^{p,q}(W)\to\La^{p+1,q}(W)$, $d_h\colon\La^{p,q}(W)\to
\La^{p,q+1}(W)$ and, moreover, as it follows from Proposition
\ref{pr:prop}~(xi), $\partial_\tau\colon \Dr^v(\La^{p,q}(W))\to
\Dr^v(\La^{p,q+1}(W))$.

\begin{remark}
The complex $(\La^q_h(W),d_h)$ is called the \emph{horizontal complex} of
the covering $\tau$, while its cohomology is the horizontal cohomology of
$\tau$. It is worth to note that $d_h$ in this case is
obtained from the de~Rham differential on the manifold $M$ by applying
the operation $\CC=\CC_\tau$ (see Remark \ref{rem:C}). From
Proposition \ref{pr:prop}~(xii) it follows that the $\CC$\h cohomology of
$\tau$ is a graded module over the graded algebra of horizontal cohomology.
\end{remark}

\section{B\"{a}cklund transformations and the main result}\label{sec:3}
Following \cite{KV-trends}, let us give a geometric definition of
B\"{a}cklund transformations. Let $\CE_i\sbs J^{k_i}(\pi_i)$, $i=1,2$, be
two differential equations and $\tau_i\colon W\to\Ei_i$ be coverings with
the same total space $W$. Then the diagram
\begin{diagram}
     &              &W&              &\\
     &\ldto^{\tau_1}& &\rdto^{\tau_2}&\\
\Ei_1&              & &              &\Ei_2\\
\end{diagram}
is called a \emph{B\"{a}cklund transformation} between the equations $\Ei_1$
and $\Ei_2$. We say that it is a \emph{B\"{a}cklund autotransformation},
if $\Ei_1=\Ei_2=\Ei$. Below we confine ourselves with autotransformations
only.

Let $\CB=(W,\tau_1,\tau_2,\CE)$ be a B\"{a}cklund autotransformation. A
point $w\in W$ is called \emph{$\tau_1$\h generic}, if the plane of the
distribution $\CC_{\tau_1}\Dr(W)$ passing through $w$ has a trivial
intersection with the tangent plane at $w$ to the fiber of $\tau_2$ passing
through the same point. Now, if $s\sbs\Ei$ is a solution of $\CE$ and
$\tau_1^{-1}(s)$ contains a $\tau_1$\h generic point, then there exists
a neighborhood $\CU$ of this point such that $\tau_2(\CU\cap\tau_1^{-1}(s))$
is fibered by solutions of $\CE$. Thus, B\"{a}cklund transformations really
determine a correspondence between solutions.

The property of a B\"{a}cklund transformation to be generic is naturally
reformulated in global terms of structural elements. Let $U_i=U_{\tau_i}$ be
the structural element of the covering $\tau_i$. Then $U_i$ may be understood
as a linear mapping $U_i\colon \Dr(W)\to \Dr(W)$, $X\mapsto X\pinner U_i$.
Moreover, $U_i$ is a projector, i.e., $U_i\circ U_i=\id$, and thus gives
the splitting
\[
\Dr(W)=\ker U_i\oplus\im U_i=\CC_{\tau_i}\Dr(W)\oplus \Dr^{v,i}(W),
\]
where $\Dr^{v,i}(W)$ is the module of $\tau_i$\h vertical vector fields on $W$.
Let us denote by
\[
U_{2,1}=\left.U_2\right|_{\Dr^{v,1}(W)}\colon \Dr^{v,1}(W)\to \Dr^{v,2}(W)
\]
the restriction of $U_2$ to $\Dr^{v,1}(W)$. Then $\CB$ is \emph{globally
$\tau_1$\h generic}, if $U_{2,1}$ is a monomorphism. It is generic in a
\emph{strong sense}, if $U_{2,1}$ is an isomorphism.

The following construction is equivalent to the definition of B\"{a}cklund
transformations. Let $\tau_i\colon W_i\to\Ei$, $i=1,2$, be two coverings
and $F\colon W_1\to W_2$ be a diffeomorphism taking the distribution
$\CC_{\tau_1}\Dr(W)$ to $\CC_{\tau_2}\Dr(W)$. Then $\CB=(W,\tau_1,
\tau_2\circ F,\CE)$ is a B\"{a}cklund transformations and any  B\"{a}cklund
transformations is formally obtained in such a way.

\begin{remark}\label{rem:triv}
It is important to stress
here that if $F$ is an isomorphism of coverings, then the B\"{a}cklund
transformation obtained in such a way is \emph{trivial} in the sense of
its action on solutions. Thus, we are interested in mappings $F$ such that
they are isomorphisms of manifolds with distributions, but not morphisms
of coverings.
\end{remark}

Assume now that a smooth family $F_\la\colon W_1\to W_2$ is given, then it
generates the corresponding family $\CB_\la$ of B\"{a}cklund transformations.
Our aim is to describe such families in sufficiently efficient terms. One
way to construct these objects is given by the following

\begin{example}[see \cite{KV-trends}]\label{ex:symm-act}
Consider an equation $\CE$, a covering $\tau\colon W\to\Ei$ over it, and
a finite symmetry $A\colon\Ei\to\Ei$. Let $\bar{A}\colon W\to W$ be a
diffeomorphic lifting of $A$ to $W$ such that
\begin{equation}\label{eq:comm}
\tau\circ\bar{A}=A\circ\tau.
\end{equation}
Denote by $\bar{A}_*\CC_\tau \Dr(W)$ the image of the distribution
$\CC_\tau \Dr(W)$ under $\bar{A}$. Then, by obvious reasons, $\bar{A}_*\CC_\tau
\Dr(W)$ determines a covering structure $U_\tau^{\bar{A}}$ in $W$ and if
$\tilde{A}$ is another lifting of $A$, then the structures
$U_\tau^{\bar{A}}$ and $U_\tau^{\tilde{A}}$ are equivalent. Thus, $\CB_A=
(W,\tau,A\circ\tau,\CE)$ is a B\"{a}cklund transformation for $\CE$.

Let $X$ be a classical infinitesimal symmetry of $\CE$ and $A_\la=\exp(\la X)
\colon\Ei\to\Ei$ the corresponding one\h parameter group of transformations
lifted to $\Ei$. Then, using the above construction, we obtain $\la$\h
parameter family of B\"{a}cklund transformations $\CB_\la=\CB_{A_\la}$.
\end{example}

\begin{remark}\label{rem:shad}
Note that since the symmetry $X$ generating the family $\CB_\la$ above
cannot be lifted as a symmetry of $W$ (i.e., as a nonlocal $\tau$\h
symmetry), it is a shadow in the covering $\tau$, as well as in all coverings
$\tau_\la=A_\la\circ\tau$.
\end{remark}

In fact, the families of B\"{a}cklund transformations obtained in the
previous example are in a sense ``counterfeit'', since, due to
\eqref{eq:comm}, their action on solutions reduces to the action of
symmetries $A_\la$. To get a ``real'' B\"{a}cklund transformation, one needs
to add into considerations an additional mapping $F\colon W\to W$
preserving the Cartan distribution on $W$ but violating \eqref{eq:comm}.

\begin{example}\label{Gordon}
Consider the infinite prolongation ${\mathcal E}^\infty$
of the sine-Gordon equation
\[
u_{xy}=\sin u
\]
and the trivial bundle
\[
\tau\colon W={\mathcal E}^\infty\otimes{\mathbb R}\to{\mathcal E}^\infty
\]
with a coordinate $v$ along fibres. Then the vector fields $D_x+X$ and
$D_t+T$, where $D_x=\CC(\partial/\partial x)$, $D_t=\CC(\partial/\partial t)$
are total derivatives and
\begin{align*}
X& = \left(-u_x + 2\la\sin\frac{u - v}2\right)\frac{\partial}{\partial v},\\
T& = \left(u_t + \frac 2\la\sin\frac{u + v}2\right)\frac{\partial}{\partial v},
\end{align*}
$\la\neq 0$, determine a one-dimensional covering structure $\tau_\la$ on the bundle
$\tau$. By changing the coordinates $u\mapsto v$, $v\mapsto u$, we get the
covering $\tau_{-\!\la}$. Denote this diffeomorphism of $W$ by $F_\la$.
Then $(W,\tau_\la,\tau_{-\!\la}\circ F_\la,{\mathcal E}^\infty)$
is the family of the classical B\"{a}cklund transformations
\begin{align*}
v_x& = -u_x + 2\la\sin\frac{u - v}2,\\
v_t& = u_t + \frac 2\la\sin\frac{u + v}2
\end{align*}
for the sine-Gordon equation. Consider the group $A_\la\colon x\mapsto\la x$,
$t\mapsto\la^{-\!1}t$ of scale symmetries of the sine-Gordon equation and
denote by ${\bar A}_\la\colon W\to W$ the diffeomorphic lifting of $A_\la$
acting trivially on the coordinate $v$. Then
\[
{\mathcal C}_{\tau_\la} \Dr(W)={\bar A}_{\la,*}({\mathcal C}_{\tau_1} \Dr(W))
\text{ and }
F_\la={\bar A}_\la\circ F_1\circ{\bar A}_{\la^{-\!1}}.
\]
\end{example}

\begin{example}\label{pKdV}
Consider now the potential KdV equation ${\mathcal E}$
\[
u_t=-u_{xxx}-3u^2_x
\]
and the trivial bundle
\[
\tau\colon W={\mathcal E}^\infty\otimes{\mathbb R}\to{\mathcal E}^\infty
\]
with a coordinate $v$ along fibres. The vector fields
\begin{align*}
X& = -\left(u_x+\frac12(v-u)^2+2\la\right)\frac{\partial}{\partial v},\\
T& = \left(u_{xxx}+u_x^2-4\la u_x-8\la^2+2u_{xx}(u-v)+(u_x-2\la)(u-v)^2\right)
\frac{\partial}{\partial v},
\end{align*}
$\la\in{\mathbb R}$, determine a one-dimensional covering structure
$\tau_\la$ on the bundle $\tau$. By changing the coordinates $u\mapsto v$,
$v\mapsto u$, we obtain the same covering $\tau_{\la}$. Denote this
diffeomorphism of $W$ by $F_\la$. Then $(W,\tau_\la,\tau_{\la}\circ
F_\la,{\mathcal E}^\infty)$ is a one-parameter family of B\"{a}cklund
transformations for the potential KdV equation
\begin{align*}
v_x& = -u_x-\frac12(v-u)^2-2\la,\\
v_t& = u_{xxx}+u_x^2-4\la u_x-8\la^2+2u_{xx}(u-v)+(u_x-2\la)(u-v)^2
\end{align*}
constructed by Wahlquist and Estabrook \cite{pKdV}. Consider the group
\begin{equation}
A_\la\colon u(x,t)\mapsto u(x-6\la t,t)+\la x-3\la^2t
\end{equation}
of symmetries of the potential KdV equation and denote by ${\bar A}_\la\colon
W\to W$ the diffeomorphic lifting of $A_\la$ acting trivially on the
coordinate $v$. Then we similarly have
\[
{\mathcal C}_{\tau_\la} \Dr(W)={\bar A}_{\la,*}({\mathcal C}_{\tau_0} \Dr(W))
\text{ and }
F_\la={\bar A}_\la\circ F_0\circ{\bar A}_{-\!\la}.
\]
\end{example}

Let us denote by
\[
\Dr^g(\La^i(W))=\{\,\Om\in \Dr^v(\La^i(W))\mid\Om(f)=0,\forall
f\in\Ci(\Ei)\,\}
\]
the module of $\tau$\h vertical derivations.
\begin{lemma}\label{lem:subcomp}
The modules $\Dr^g(\La^i(W))$ are invariant with respect to the differential
$\partial_\tau$\textup{:}
\[
\partial_\tau\big(\Dr^g(\La^i(W))\big)\sbs \Dr^g(\La^{i+1}(W)).
\]
\end{lemma}
\begin{proof}
Let $\Om\in \Dr^g(\La^i(W))$ and $f\in\Ci(\Ei)$. Then due to the definition
of the Fr\"{o}licher--Nijenhuis bracket one has
\[
(\partial_\tau(\Om))(f)=\fnij{U_\tau}{\Om}(f)
=\Ld_{U_\tau}(\Om(f))-(-1)^\Om \Ld_\Om(U_\tau(f)).
\]
The first summand vanishes, since $\Om\in \Dr^g(\La^i(W))$. On the other hand,
$U_\tau(f)=U_\CE(f)$ and consequently is a one-form on $\Ei$. Hence, the
second summand vanishes as well.
\end{proof}

Denote by $\partial_g\colon \Dr^g(\La^i(W))\to \Dr^g(\La^{i+1}(W))$ the
restriction of $\partial_\tau$ to $\Dr^g(\La^i(W))$ and by
\[
\partial_s\colon \Dr^s(\La^i(W))\to \Dr^s(\La^{i+1}(W))
\]
the corresponding quotient complex, where, by definition,
\[
\Dr^s(\La^i(W))=\Dr^v(\La^i(W))/\Dr^g(\La^i(W)).
\]
Then the short exact sequence of complexes
\begin{diagram}[width=1em]
 &    &0     &                    &0            &    &     &    &0            &                    &0                &    &     \\
 &    &\dto  &                    &\dto         &    &     &    &\dto         &                    &\dto             &    &     \\
0&\rto&\Dr^g(W)&\rto^{\partial_g}   &\Dr^g(\La^1(W))&\rto&\dots&\rto&\Dr^g(\La^i(W))&\rto^{\partial_g}   &\Dr^g(\La^{i+1}(W))&\rto&\dotsb\\
 &    &\dto  &                    &\dto         &    &     &    &\dto         &                    &\dto             &    &     \\
0&\rto&\Dr^v(W)&\rto^{\partial_\tau}&\Dr^v(\La^1(W))&\rto&\dots&\rto&\Dr^v(\La^i(W))&\rto^{\partial_\tau}&\Dr^v(\La^{i+1}(W))&\rto&\dotsb\\
 &    &\dto  &                    &\dto         &    &     &    &\dto         &                    &\dto             &    &     \\
0&\rto&\Dr^s(W)&\rto^{\partial_s}   &\Dr^s(\La^1(W))&\rto&\dots&\rto&\Dr^s(\La^i(W))&\rto^{\partial_s}   &\Dr^s(\La^{i+1}(W))&\rto&\dotsb\\
 &    &\dto  &                    &\dto         &    &     &    &\dto         &                    &\dto             &    &     \\
 &    &0     &                    &0            &    &     &    &0            &                    &0                &    &     \\
\end{diagram}
is defined.

Denote by $H_g^i(\CE;\tau)$ and $H_s^i(\CE;\tau)$ the cohomology of the
top and bottom lines respectively. Then one has the long exact cohomology
sequence
\begin{multline}\label{eq:exact}
0\to H_g^0(\CE;\tau)\to H_\CC^0(\CE;\tau)\to H_s^0(\CE;\tau)
\xrightarrow{\phi}H_g^1(\CE;\tau)\to H_\CC^1(\CE;\tau)\to H_s^1(\CE;\tau)\to\\
\dotsb\to H_g^i(\CE;\tau)\to H_\CC^i(\CE;\tau)\to H_s^i(\CE;\tau)\to\dotsb,
\end{multline}
where $\phi$ is the connecting homomorphism.

Similar to Theorem~\ref{thm:deform}, we have the following result:
\begin{proposition}\label{pr:deform}
In the situation above one has\textup{:}
\begin{enumerate}
\item
The module $H_g^0(\CE;\tau)$ consists of \textup{``}gauge\textup{''}
symmetries in the covering $\tau$\textup{,} i.e.\textup{,} of nonlocal
$\tau$\h symmetries vertical with respect to the projection $\tau$.
\item
The module $H_s^0(\CE;\tau)$ coincides with the set of $\tau$\h shadows in
the covering $\tau$.
\item
The module $H_g^1(\CE;\tau)$ consists of equivalence classes of deformations
of the covering structure $U_\tau$ acting trivially on the equation structure
$U_\CE$.
\end{enumerate}
\end{proposition}

Now, combining the last result with exact sequence \eqref{eq:exact}, we
obtain the following fundamental theorem:
\begin{theorem}\label{fundamental}
Let $\tau\colon W\to\CE$ be a covering and $A_\la\colon W\to W$ be a smooth
family of diffeomorphisms such that $A_0=\id$ and $\tau_\la=\tau\circ
A_\la\colon W\to\CE$ is a covering for any $\la\in{\mathbb R}$. Then
$U_{\tau_\la}$ is of the form
\begin{equation}\label{eq:infin}
U_{\tau_\la}=U_\tau+\la\fnij{U_\tau}{X}+O(\la^2),
\end{equation}
where $X$ is a $\tau$\h shadow\textup{,} i.e.\textup{,} all smooth families
corresponding to the covering $\tau$ are infinitesimally identified
with $\im\partial_s$.
\end{theorem}
\begin{proof}
The family of coverings $\tau_\la$
is a deformation of $\tau$. Since we work with deformations which leave
the equation structure unchanged, then, by Proposition \ref{pr:deform},
their infinitesimal parts are elements of $H_g^1(\CE;\tau)$. Let $\Om$
be such an element.

Now, by Remark \ref{rem:triv}, the deformation we are dealing with is
to be trivial as a deformation of $W$ endowed with the structure $U_\tau$.
On the infinitesimal level, this means that the image of $\Om$ in
$H^1(\CE;\tau)$ should vanish. But by exactness of \eqref{eq:exact} we see
that $\Om=\phi(X)$ for some $X\in H_s^0(\CE;\tau)$. It now suffices to
note that by construction of the connecting homomorphism, $\phi(X)=
\fnij{U_\tau}{X}$.

The family $A_\la$ allows us to find a shadow $X$ explicitly. Namely,
we obviously have
\begin{multline*}
\left.\frac{d}{d\la}\right|_{\la=0}U_{\tau_\la}=
\left.\frac{d}{d\la}\right|_{\la=0}A_{\la,*}(\Ld_{U_\tau})=\\
=\left.\frac{d}{d\la}\right|_{\la=0}
A_\la^*\circ \Ld_{U_\tau}\circ(A_\la^*)^{-1}=[\Ld_Y,\Ld_{U_\tau}]=
\Ld_{\fnij{Y}{U_\tau}},
\end{multline*}
where
\[
Y=\left.\frac{dA_\la}{d\la}\right|_{\la=0}\in \Dr(W).
\]
Hence, infinitesimal action is given by the Fr\"olicher\,--\,Nijenhuis
bracket. In the coset
$X\bmod\CC_\tau \Dr(W)\in\sym_\tau\CE$ there exists a unique
$(\pi_\CE\circ\tau)$\h vertical representative $X$, and the corresponding
element $[X]\in H_s^0(\CE;\tau)$ is a required shadow.
\end{proof}

\begin{remark}
Consider the one-parameter families of coverings $\tau_\la$ and $\tau'_\la$
from Examples \ref{Gordon} and \ref{pKdV} respectively. The classical
infinitesimal symmetries corresponding to the one-parameter groups
$A_\la$ of finite symmetries are
\begin{align*}
x\frac{\partial}{\partial x}&-t\frac{\partial}{\partial t}&&
\text{ for the sine-Gordon equation},\\
x\frac{\partial}{\partial u}&-6t\frac{\partial}{\partial x}&&
\text{ for the potential KdV equation}.
\end{align*}
The corresponding higher symmetries are the shadows of $\tau_1$ and $\tau'_0$
respectively (see Example \ref{sym-shad}). These shadows determine the
infinitesimal parts of the families $U_{\tau_\la}$ and $U_{\tau'_\la}$
according to Theorem \ref{fundamental}.
\end{remark}
\begin{remark}
Denote by ${\mathcal C}ov(\tau)$ the ``manifold'' of all coverings
obtained from the covering $\tau$ by the above described way. Then from
exactness of \eqref{eq:exact} it follows that the tangent plane to
${\mathcal C}ov(\tau)$ at $\tau$ is identified with the space
$\mathrm{shad}_\tau\CE/
\overline{\sym}_\tau\CE$, where $\mathrm{shad}_\tau\CE=H_s^0(\CE;\tau)$ is
the space of all $\tau$\h shadows. Finally, the space
$\overline{\sym}_\tau\CE=
\sym_\tau\CE/\sym_\tau^g\CE$ is the quotient of all $\tau$\h symmetries
over gauge ones.
\end{remark}

\section*{Acknowledgments}
It is our pleasant obligation to express deep gratitude to Michal Marvan
for a long and fruitful collaboration in nonlocal aspects of PDE geometry.


\end{document}